\documentclass[12pt]{article}

\usepackage{graphicx}

\newcommand{\tr}{\mbox{Tr} \, }
\newcommand{\ket}[1]{\left | #1 \right \rangle}
\newcommand{\bra}[1]{\left \langle #1 \right |}

\newcommand{\proj}[1]{\ket{#1} \! \bra{#1}}
\newcommand{\outerprod}[2]{\ket{#1} \! \bra{#2}}

\newcommand{\superop}{{\cal E}}
\newcommand{\unity}{\mbox{\bf I}}
\newcommand{\hilbert}{{\cal H}}

\newcommand{\bigsum}[1]{{\displaystyle \sum_{#1}}}

\begin{document}

\title{Quantum mutual information and the one-time pad}

\author{Benjamin Schumacher$^{(1)}$
        and Michael D. Westmoreland$^{(2)}$}
\maketitle
\begin{center}
{\sl
$^{(1)}$Department of Physics, Kenyon College, Gambier, OH 43022 USA \\
$^{(2)}$Department of Mathematical Sciences, Denison University,
 Granville, OH  43023 USA }
\end{center}

\section*{Abstract}

Alice and Bob share a correlated composite quantum system $AB$.  If
$AB$ is used as the key for a one-time pad cryptographic system, we
show that the maximum amount of information that Alice 
can send securely to Bob is the quantum mutual information 
of $AB$.

\section{The one-time pad and mutual information}

A {\em one-time pad} \cite{onetimepad} 
is a cryptographic protocol in which communicators 
Alice and Bob initially have correlated random variables, collectively 
called the ``key'', that are not correlated with any variables possessed 
by a potential eavesdropper Eve.  (In most discussions, the key 
variables possessed by Alice and Bob are supposed to be perfectly 
correlated---e.g., they are identical copies of the same secret string 
of bits.  We consider the more general case.)  If the key variables are 
used only once, they allow Alice to send Bob a perfectly secret 
message over a public communication channel.  The value of a key
as a resource is the amount of information that can be sent secretly 
by its use.

In this paper we examine a quantum mechanical analogue of the one-time
pad.  Alice and Bob initially share a correlated composite quantum system
$AB$.  Alice encodes a classical message by performing one of several
possible operations on her subsystem $A$, after which she transfers it
to Bob.  Bob reads the message via a measurement on the entire system
$AB$.  The eavesdropper Eve only has access to subsystem $A$; thus,
to ensure the security of the secret message, Alice must ensure that
the $A$ by itself can provide no information to Eve.

Holevo \cite{holevo1}
provided an upper bound for the accessible information in a 
measurement.  Suppose a quantum system is prepared in a state
$\rho_{\alpha}$ with probability $p_{\alpha}$.  The ensemble average
state is $\rho = \bigsum{\alpha} p_{\alpha} \rho_{\alpha}$.  Holevo
showed that, for any measurement, the mutual information $I$ 
between the preparation and the measurement result is bounded 
above by
\begin{equation}
    I \leq \chi = S \left ( \rho \right ) -
        \sum_{\alpha} p_{\alpha} \, S \left ( \rho_{\alpha} \right ),
\end{equation}
where $S(\rho) = - \tr \rho \log \rho$.  Holevo \cite{holevo2} 
and Schumacher and Westmoreland \cite{noisy}
proved that, with appropriate choices of code and 
decoding observable, this upper bound can be approached asymptotically. 
Therefore, $\chi$ measures the classical information that can be 
conveyed using a particular ensemble of quantum states.

The quantity $\chi \geq 0$, with equality if and only if all of the
possible states $\rho_{\alpha}$ are the same.  We can say even more.
The {\em only} situation in which zero information is provided by
any measurement is the situation in which all of the possible states
are the same.  Since Alice wishes to exclude the eavesdropper, she
must arrange that her various operations always lead to the same
output state of $A$.  That is, $\chi^{A} = 0$.

However, Alice and Bob want to make sure that $\chi^{AB} > 0$, since
Bob needs to read the secret message by an $AB$ measurement.  Let
$\rho^{AB}$ be the initial ``key'' state of $AB$.  Only the correlations 
within $\rho^{AB}$ permit Alice and Bob to communicate at all.  
If the initial state $\rho^{AB}$ is a product state, then
it must remain a product state regardless of Alice's manipulation of
it---and always the same product state, since $\rho^{B}$ is unchanged
and Alice's final state $\sigma^{A}$ is fixed.  Even with both 
$A$ and $B$ in his possession, Bob will not be able to infer 
anything about Alice's choice of operation, because he will always
have the state $\sigma^{A} \otimes \rho^{B}$.  Without correlations,
the ``key'' state $\rho^{AB}$ is useless. 

We now put this intuitive observation on a more quantitative basis.
Imagine that Alice performs the operation $\superop^{A}_{\alpha}$ on $A$ with
probability $p_{\alpha}$.  We write
\begin{eqnarray}
    \sigma^{AB}_{\alpha} & = & \left ( \superop^{A}_{\alpha} \otimes \unity^{B} 
        \right ) \rho^{AB} \\
    \sigma^{AB} & = & \sum_{\alpha} p_{\alpha} \, \sigma^{AB}_{\alpha} .
\end{eqnarray}
To exclude the eavesdropper, we require that $\sigma^{A}_{\alpha} =
\sigma^{A}$ for every $\alpha$.
The information that Alice can send to Bob will be limited by
\begin{equation}
    \chi^{AB} = S \left ( \sigma^{AB} \right )  - 
                \sum_{\alpha} p_{\alpha} \, S \left ( \sigma^{AB}_{\alpha} \right ) .
\end{equation}

The entropy of the average state $\sigma^{AB}$ is subadditive, so
that $S \left ( \sigma^{AB} \right ) \leq S \left ( \sigma^{A} \right )
+ S \left ( \sigma^{B} \right )$ (with equality if and only if
$\sigma^{AB} = \sigma^{A} \otimes \sigma^{B}$).
Thus,
\begin{equation}
    \chi^{AB} \leq S \left ( \sigma^{B} \right ) 
                   + S \left ( \sigma^{A} \right )
                   - \sum_{\alpha} p_{\alpha} \, S \left ( \sigma^{AB}_{\alpha} \right ).
\end{equation}
Note that $\sigma^{B} = \rho^{B}$ (since Alice only operates on $A$)
and that, by assumption, the individual final $A$ states satisfy 
$\sigma^{A}_{\alpha} = \sigma^{A}$ for all $\alpha$:
\begin{equation}
    \chi^{AB} \leq S \left ( \rho^{B} \right ) +
                \sum_{\alpha} p_{\alpha} \left ( 
                S \left ( \sigma^{A}_{\alpha} \right ) 
                    - S \left ( \sigma^{AB}_{\alpha} \right ) \right ).
\end{equation}
No operation on $A$ alone can lead to an increase in the 
coherent information \cite{coherent}
$S^{A} - S^{AB}$, so that for all $\alpha$,
\begin{equation}
   S \left ( \sigma^{A}_{\alpha} \right ) 
                    - S \left ( \sigma^{AB}_{\alpha} \right ) \leq
   S \left ( \rho^{A} \right ) 
                    - S \left ( \rho^{AB} \right ) .  \label{coherent-bound}
\end{equation}
Therefore,
\begin{equation}
    \chi^{AB} \leq S \left ( \rho^{A} \right ) +
                    S \left ( \rho^{B} \right ) -
                    S \left ( \rho^{AB} \right ) .
\end{equation}
The quantity on the right is $I_{\rho}(A:B)$, the quantum mutual 
information between $A$ and $B$, a measure of the degree 
of correlation in the original state $\rho^{AB}$.
We have shown that the information that Alice can transmit secretly 
to Bob using $\rho^{AB}$ as a one-time pad is bounded above
by $I_{\rho}(A:B)$.

\section{A special case}

Having shown that $\chi^{AB} \leq I_{\rho}(A:B)$, we will now
show that Alice can choose an ensemble of operations so that 
$\chi^{AB} \rightarrow I_{\rho}(A:B)$ asymptotically.
Since we know that we can achieve $\chi^{AB}$ as an 
asymptotic information rate, it follows that Alice can 
send up to $I_{\rho}(A:B)$ bits per key to Bob while keeping
Eve completely excluded.

To do this, we will only need to consider unitary
operations on $A$, given by unitary operators $U^{A}_{\alpha}$.
The new $A$ states will be exactly the same as the original
``key'' state of $A$, so that
\begin{equation}
\sigma_{\alpha}^{A} = U_{\alpha} \rho^{A} {U_{\alpha}}^{\dagger} 
                    = \rho^{A}
\end{equation}
for all $\alpha$.  This amounts to saying that 
$\left [ U^{A}_{\alpha} , \rho^{A} \right ] = 0$.

We will first consider a special case in which we can make
$\chi^{AB} = I_{\rho}(A:B)$ in a single composite system, without the
need for an asymptotic argument.  Suppose that the initial
$A$ state is maximally mixed on a subspace, so that 
$\rho^{A} = \frac{1}{d} \Pi$ (where $\Pi$ is the projection onto
a $d$-dimensional subspace).  Then any unitary operator on $A$ 
that commutes with $\Pi$ will leave $\rho^{A}$ invariant.  
Let us choose basis states $\ket{k^{A}}$ for the support of $\Pi$ 
and write
\begin{equation}
    \rho^{AB} = \sum_{kl} \outerprod{k^{A}}{l^{A}} \otimes w^{B}_{kl} .
\end{equation}
By considering $\rho^{B} = \tr_{A} \rho^{AB}$, we can see that the
$B$ operators $w^{B}_{kl}$ satisfy
\begin{equation}
    \rho^{B} = \sum_{k} w^{B}_{kk} .
\end{equation}

What operators $U^{A}_{\alpha}$ does Alice include in her ensemble?
We will say that her ensemble includes
\begin{itemize}
    \item  All possible relative phase flips among 
        the $\ket{k^{A}}$ basis states;
    \item  All permutations of the $\ket{k^{A}}$ basis states; and
    \item  All combinations of these.
\end{itemize}
There are $N$ such operators, and Alice uses each with probability
$1/N$.  Thus,
\begin{eqnarray}
    \sigma^{AB}_{\alpha} & = & \sum_{kl} \left ( U^{A}_{\alpha} 
                        \outerprod{k^{A}}{l^{A}} 
                        {U^{A}_{\alpha}}^{\dagger} \right )
                        \otimes w^{B}_{kl} \\
    \sigma^{AB} & = & \sum_{kl}  \left ( \frac{1}{N} \sum_{\alpha} 
                        U^{A}_{\alpha} 
                        \outerprod{k^{A}}{l^{A}} 
                        {U^{A}_{\alpha}}^{\dagger} \right )
                        \otimes w^{B}_{kl} .
\end{eqnarray}
Consider the sum in the second expression.  When $k \neq l$, the sum
over $\alpha$ contains all relative phase flips among the $A$ basis
states with equal weights.  In this case the sum must equal zero.
The expression for $\sigma^{AB}$
becomes
\begin{equation}
    \sigma^{AB} = \sum_{k}  \left ( \frac{1}{N} \sum_{\alpha} 
                        U^{A}_{\alpha} 
                        \outerprod{k^{A}}{k^{A}} 
                        {U^{A}_{\alpha}}^{\dagger} \right )
                        \otimes w^{B}_{kk} .
\end{equation}
The sum over $\alpha$ also includes all permutations among 
the $A$ basis states.  This means that the result of this sum 
is independent of $k$.  We conclude that the average state
$\sigma^{AB}$ is a product state, namely
\begin{equation}
    \sigma^{AB} = \rho^{A} \otimes \rho^{B} .
\end{equation}

For each $\alpha$, $\sigma^{AB}_{\alpha}$ is just the original
state $\rho^{AB}$, rotated by the unitary operator $U^{A}_{\alpha}$.
This rotated state will have the same entropy as the original.
It follows that 
\begin{eqnarray}
    \chi^{AB} & = & S \left ( \sigma^{AB} \right )  - 
        \sum_{\alpha} p_{\alpha} \, S \left ( \sigma^{AB}_{\alpha} \right )
        \nonumber \\
        & = & S \left ( \rho^{A} \right ) + S \left ( \rho^{B} \right )
                - S \left ( \rho^{AB} \right )  \\
        & = & I_{\rho}(A:B) .
\end{eqnarray}
In this special case, then, we can arrange for $\chi^{AB}$ to achieve
its upper bound of $I_{\rho}(A:B)$ exactly.

Notice how this works.  We have arranged Alice's ensemble of operations
so that the correlations between $A$ and $B$ completely disappear on
average---leaving $\sigma^{AB}$ a product state.  Let us think about this
more generally.  Once again, we suppose that we have a bunch of unitary
operators $U^{A}_{\alpha}$ acting on $A$, which do not alter the subsystem
state $\rho^{A}$.  We have
\begin{equation}
    \chi^{AB} = S \left ( \sigma^{AB} \right ) - S \left ( \rho^{AB} \right )
\end{equation}
(since for each $\alpha$ the state $\sigma^{AB}_{\alpha}$ has the same
entropy as $\rho^{AB}$).  Noting that $\sigma^{A} = \rho^{A}$ and
$\sigma^{B} = \rho^{B}$, we can rewrite this as
\begin{equation}
    \chi^{AB} = I_{\rho}(A:B) - I_{\sigma}(A:B) ,  \label{mutual-reduction}
\end{equation}
where $I_{\rho}(A:B)$ and $I_{\sigma}(A:B)$ are the mutual informations for
$\rho^{AB}$ and $\sigma^{AB}$, respectively.  In other words, $\chi^{AB}$ 
is exactly the amount by which we have, on average, reduced the mutual
information between the systems.  In our special case, where the subsystem
$A$ is completely mixed, we can reduce this all the way to zero, and so
$\chi^{AB} = I_{\rho}(A:B)$.

This points up a connection between our analysis and the work of 
Groisman et al. \cite{groisman}
, who define the ``total correlation'' of two systems 
to be the amount of classical information that must be added to the
system so that the correlations can be completely eliminated by local
operations.  They show that the total correlation is given by the
quantum mutual information.  The elimination of correlations is not
our aim; rather, we wish to maximize $\chi^{AB}$ subject to the strict 
privacy condition that $\chi^{A} = 0$.  Nevertheless, 
Equation~\ref{mutual-reduction} tells us that these two tasks
are closely related.

\section{The general case}

Now let us consider a general state $\rho^{AB}$.  The subsystem state
$\rho^{A}$ has $D$ distinct eigenvalues $\lambda_K$.  For a given
$K$, the eigenspace of $\lambda_{K}$ has dimension $d_{k}$.  We
can therefore choose a basis of $\rho^{A}$ eigenstates and write
\begin{equation}
    \rho^{A} = \sum_{K=1}^{D} \lambda_{K} \left ( \sum_{m_K = 1}^{d_K} 
                \proj{K m_{K}} \right ).  \label{rhoA-expression}
\end{equation}
For a given $K$, we think of the basis states $\ket{K m_{K}}$ as
comprising a ``block'' spanning the $d_K$-dimensional eigenspace
of $\lambda_{K}$.  This block has total ``weight'' $P_K = d_K \lambda_K$
in this mixture.  We can write
\begin{equation}
    \rho^{A} = \sum_{K} P_{K} \, \rho^{A}_{K} ,
\end{equation}
where each of the $\rho^{A}_{K}$ is the density operator that is
maximally mixed on the eigenspace of $\lambda_{K}$:
\begin{equation}
    \rho^{A}_{K} = \sum_{m_K} \frac{1}{d_K} \proj{K m_K} .
\end{equation}

The joint state $\rho^{AB}$ can be written
\begin{equation}
    \rho^{AB} = \sum_{KL} \left (  \sum_{m_K n_L} 
                \outerprod{K m_K}{L n_L} \otimes w^{B}_{K m_K L n_L}
                \right ) .
\end{equation}
What can we say about the operators $w^{B}_{K m_K L n_L}$?  
If we compare the partial trace of this expression with 
Equation~\ref{rhoA-expression}, we see that
\begin{equation}
    \tr w^{B}_{K m_K L n_L}  =  \lambda_{K} \, \, \delta_{KL}
            \,  \delta_{m_K n_L} .
\end{equation}
Thus, given a value of $K$,
\begin{equation}
    \sum_{m_K} w^{B}_{K m_K K m_K} = P_{K} \rho^{B}_{K}
\end{equation}
for some density operator $\rho^{B}_{K}$.  This will be useful below.

Notice that, for various values of $K$, the density operators $\rho^{A}_{K}$
have orthogonal supports.  In general, we can make no such claim about the
supports of the density operators $\rho^{B}_{K}$.

As before, Alice will perform unitary operations on $A$ that do not 
change the subsystem state $\rho^{A}$.  The operators $U^{A}_{\alpha}$
include
\begin{itemize}
    \item  All relative phase flips between distinct blocks;
    \item  All relative phase flips between basis states within
            each block;
    \item  All permutations of the basis states within each block;
            and
    \item  All combinations of these.
\end{itemize}
Again, we say that there are $N$ such operators, and Alice uses each
with probability $1/N$.  

The resulting average state $\sigma^{AB}$ is
\begin{equation}
    \sigma^{AB} = \sum_{KL} \sum_{m_K n_L} \left ( \frac{1}{N}
                \sum_{\alpha} U^{A}_{\alpha} \outerprod{K m_K}{L n_L} 
                {U^{A}_{\alpha}}^{\dagger} \right )
                \otimes w^{B}_{K m_K L n_L} .
\end{equation}
Since the average over $\alpha$ includes all phase flips between
distinct values of $K$ and $L$, the average in parentheses is zero
unless $K = L$, so
\begin{equation}
    \sigma^{AB} = \sum_{K} \sum_{m_K n_K} \left ( \frac{1}{N}
                \sum_{\alpha} U^{A}_{\alpha} \outerprod{K m_K}{K n_K} 
                {U^{A}_{\alpha}}^{\dagger} \right )
                \otimes w^{B}_{K m_K K n_K} .
\end{equation}
Also, we include all phase flips between distinct values of $m_K$
and $n_K$, so the sum becomes
\begin{equation}
    \sigma^{AB} = \sum_{K \, m_K} \left ( \frac{1}{N}
                \sum_{\alpha} U^{A}_{\alpha} \outerprod{K m_K}{K m_K} 
                {U^{A}_{\alpha}}^{\dagger} \right ) 
                \otimes w^{B}_{K m_K K m_K} .
\end{equation}
Finally, since the $U^{A}_{\alpha}$ operators include all permutations
of basis states within a given block, the average in parenthesis 
depends only on $K$ and not on $m_K$.  Indeed, this average is the
uniform density operator on the $\lambda_{K}$-eigenspace for $\rho^{A}$,
which is just $\rho^{A}_{K}$.  This means we can write
\begin{equation}
    \sigma^{AB} = \sum_{K} P_{K} \, \rho^{A}_{K} \otimes \rho^{B}_{K} .
\end{equation}

From this, noting that the $\rho^{A}_{K}$ operators have orthogonal supports,
we can calculate the quantum mutual information $I_{\sigma}(A:B)$ to be
\begin{equation}
    I_{\sigma}(A:B) = S \left ( \rho^{B} \right ) 
                    - \sum_{K} P_{K} S \left ( \rho^{B}_{K} \right ).
\end{equation}
The right-hand side of this equation is bounded above by $\log D$, the
logarithm of the number of distinct eigenvalues of $\rho^{A}$ (and thus
the number of values of the eigenvalue index $K$).  Therefore,
\begin{equation}
    I_{\sigma}(A:B) \leq \log D .
\end{equation}
Alice can therefore achieve a Holevo bound for the composite system 
satisfying
\begin{equation}
    \chi^{AB} \geq I_{\rho}(A:B) - \log D .
\end{equation}

Now consider the asymptotic problem.  Alice and Bob share a large
number $n$ of copies of the pair $AB$, so that their initial joint
state is $\left ( \rho^{AB} \right )^{\otimes n}$.  The quantum
mutual information of this state is just $n \, I_{\rho}(A:B)$.
Alice performs operations on all of her copies together such
that the final state of these copies is always the same.  Alice's
systems are delivered to Bob, who will try to distinguish which
operation Alice performed.  Regardless of Alice's operations,
\begin{equation}
    \frac{1}{n} \, \chi^{(AB)^{\otimes n}} \leq I_{\rho}(A:B) .
\end{equation}
We will now show that, for a suitable ensemble of operations,
Alice can approach equality, and therefore $I_{\rho}(A:B)$ is an
asymptotically achievable information rate from Alice to Bob
as $n \rightarrow \infty$.

First, we note that $\left ( \rho^{A} \right )^{\otimes n}$ is a
highly degenerate state for large $n$.  If the Hilbert space 
$\hilbert^{A}$ has dimension $d$, then $(\hilbert^{A})^{\otimes n}$
has dimension $d^{n}$ (exponential in $n$), but the state 
$\left ( \rho^{A} \right )^{\otimes n}$ has no more than 
$(n+1)^{d}$ (polynomial in $n$) distinct eigenvalues.  These
distinct eigenvalues correspond to the {\em type classes}
\cite{cover} of
sequences of $n$ i.i.d. random variables, each having $d$
values.  Therefore, if we use our previous method to choose
an ensemble of unitary operators for Alice's systems that each
leave $\left ( \rho^{A} \right )^{\otimes n}$ unchanged, we can
create an ensemble of $(AB)^{\otimes n}$ states such that
\begin{equation}
    \chi^{(AB)^{\otimes n}} \geq n I_{\rho}(A:B) - \log (n+1)^{d} .
\end{equation}
Therefore,
\begin{equation}
    \frac{1}{n} \, \chi^{(AB)^{\otimes n}}  \geq I_{\rho}(A:B) 
            - \frac{d}{n} \log (n+1) .
\end{equation}
Since the second term goes to zero as $n \rightarrow \infty$, 
we have found a sequence of procedures such that
\begin{equation}
    \lim_{n \rightarrow \infty} \frac{1}{n} \, \chi^{(AB)^{\otimes n}}
        = I_{\rho}(A:B) .
\end{equation}
The mutual information $I_{\rho}(A:B)$ is therefore the information
capacity from Alice to Bob if Alice can perform only local operations
on the $A$ systems that always lead to the same $A$ state (and will
thus completely exclude any eavesdropper with access only to $A$).

\section{Slightly insecure}

Note that we have required absolute perfection---that is, we have required
that, by examining system $A$ by itself, the eavesdropper Eve cannot get any 
information at all.  No matter what operation Alice performs, the final 
$A$ state is exactly the same.  But what if we relax this requirement?
Since Alice now has a wider range of operations at her disposal, she 
should be able to increase the Holevo bound $\chi^{AB}$, and thus the
information that she can deliver to Bob.  If Eve has access only to a
finite specified amount of information, how much additional capacity can
Alice and Bob achieve?
We will now show that the extra capacity from Alice
to Bob is no larger than the Holevo bound $\chi^{A}$, 
which in turn bounds the accessible information of the eavesdropper.
Thus, if the protocol is only slightly insecure ($\chi^{A}$ is
small), the information capacity is only slightly increased.

We begin with the key state $\rho^{AB}$, and Alice performs the operation
$\superop^{A}_{\alpha}$ on $A$ with probability $p_{\alpha}$.  We do not
require the operations to be unitary.  As before, the final
states are
\begin{eqnarray}
    \sigma^{A}_{\alpha} & = & \superop^{A}_{\alpha}
                            \left ( \rho^{A} \right ) \\
    \sigma^{AB}_{\alpha} & = & \superop^{A}_{\alpha} \otimes \unity^{B}
                            \left ( \rho^{AB} \right ) \\
    \sigma^{A} & = & \sum_{\alpha} p_{\alpha} \, \sigma^{A}_{\alpha} \\
    \sigma^{AB} & = & \sum_{\alpha} p_{\alpha} \, \sigma^{AB}_{\alpha} .
\end{eqnarray}
Then
\begin{equation}
    \chi^{AB} - \chi^{A}  =  S \left ( \sigma^{AB} \right )
                                - S \left ( \sigma^{A} \right )
                                + \sum_{\alpha} p_{\alpha} \, \left (
                                S \left ( \rho^{A}_{\alpha} \right )
                                - S \left ( \rho^{AB}_{\alpha} \right )
                                \right ) .
\end{equation}
By Equation~\ref{coherent-bound}, remembering that
$S \left ( \sigma^{B} \right ) = S \left ( \rho^{B} \right )$ for
$A$ operations, this becomes
\begin{eqnarray}
    \chi^{AB} - \chi^{A} & \leq & S \left ( \sigma^{AB} \right ) -
                            S \left ( \sigma^{A} \right ) +
                            S \left ( \rho^{A} \right ) -
                            S \left ( \rho^{AB} \right )  \\
            & = & I_{\rho}(A:B) - I_{\sigma}(A:B) \\
            & \leq & I_{\rho}(A:B) .
\end{eqnarray}
Thus,
\begin{equation}
    \chi^{AB} \leq I_{\rho}(A:B) + \chi^{A} .
\end{equation}
Allowing a small non-zero $\chi^{A}$ can only increase $\chi^{AB}$ by that
same small amount.

\section{Remarks}

In our analysis of the quantum problem, we have also proven the
analogous classical result.  That is, suppose Alice and Bob possess 
a pair of correlated random variables $X_A$ and $X_B$.  Alice encodes
her message by performing one of several possible operations on her 
own variable $X_A$.  To prevent Eve (who has access to $X_A$) from
reading the message, she arranges for the marginal 
probability distribution of $X_A$ to be independent of her message.
Bob receives $X_A$ and reads the message by examining the joint
value $\left ( X_A , X_B \right )$.  In such a situation, 
the maximum achievable secure communication rate from Alice to Bob 
is the classical mutual information $I\left( X_A : X_B \right )$.  
This follows from our quantum result in the case that the quantum 
state of the composite system $AB$ is a mixture of products of states 
drawn from orthogonal sets for $A$ and $B$.

In other words, our analysis tells us that the mutual information 
is the answer to the {\em same} communication problem in both the 
classical and quantum settings.  This illuminates the connections 
between classical and quantum information ideas.  In particular, 
it sheds light on the meaning of the mutual information
functional as a measure of the degree of correlation between 
physical systems.

We would like to thank A. Winter for valuable suggestions.  We 
also acknowledge helpful discussions of this work with the 
Kenyon-Denison quantum  information research group, including 
M. Nathanson, L. Kennard and K. Christandl.

\end{document}